\documentclass[preprint,12pt,3p]{elsarticle}
\usepackage{graphicx,pslatex,float,color,array,amssymb,amsmath,euscript}
\usepackage{subcaption}
\usepackage{lipsum}
\usepackage{siunitx}
\usepackage{xcolor}
\usepackage{graphicx}
\usepackage{lipsum} 
\usepackage{dblfloatfix}
\usepackage[position=b]{subcaption}
\usepackage{graphicx,pslatex,float,color,array,amssymb,amsmath,euscript}
\usepackage{subcaption}
\usepackage{lipsum}
\usepackage{siunitx}
\usepackage{multirow}
\biboptions{sort&compress}

\def\ten#1{\oalign{$\bf #1$\crcr\hidewidth$\scriptscriptstyle\sim$\hidewidth}\vphantom{#1}}

\def\ordq#1{\ten{\ten#1}}

\def\vect#1{{\bf \underline #1}{\vphantom{#1}}}
\def\nab{\vect \nabla}

\definecolor{Blue}{rgb}{0.3,0.3,0.9}
\definecolor{Std}{rgb}{0.,0.,0.}
\def\rev#1{\color{Std}{#1}\color{Std}}

\usepackage{caption}
\journal{Superalloys 2024, proceedings}
\begin{document}

\title{Effect of free surface, oxide and coating layers on rafting in $\gamma-\gamma'$ superalloys}
\author[add2]{Wajih Jbara}
\author[add2,cor]{Vincent Maurel}
\author[add2]{Kais Ammar}
\author[add2]{Samuel Forest}
\cortext[cor]{Corresponding author. E-mail address: vincent.maurel@minesparis.psl.eu (V. Maurel)}
\address[add2]{MINES Paris, PSL University, MAT - Centre des Matériaux, CNRS UMR 7633, BP 87 91003 Evry, France}
\begin{abstract}
    Complex microstructure evolution has been observed \rev{both bare and coated } Ni-based single crystal superalloys. Rafting and $\gamma'$ depletion are investigated in this study through a brief experimental analysis and a detailed phase field model to account for mechanical-diffusion coupling. The proposed model has been implemented in a finite element code. As a main result, it is shown that rafting, $\gamma'$ depletion close to free surface/oxide layer or $\gamma'$ coalescence close to coating layer, and mechanical behavior are strongly coupled. The local additional flux of Al explains this coupling to a large extent. Finally, a discussion of strain localization and local flux of Al paves the way for clarification of these cases that degrade the performance of superalloys.
\end{abstract}

\maketitle



\section{Introduction}



Creep behavior in superalloys is a critical property that is influenced by factors such as temperature range and stress level, and affects \rev{both } microstructure evolution and final mechanical properties \cite{reed2008}. For single crystal Ni-based superalloys, optimization relies on achieving optimal volume fraction, shape and size of $\gamma'$ precipitates in a coherent $\gamma$ \rev{matrix } to enhance the precipitate hardening mechanism \cite{khan1985superalliages,murakumo2004creep}. In high-temperature applications such as high-pressure turbine blades and vanes, understanding the driving forces for microstructure evolution under severe conditions is paramount \cite{mughrabi2009microstructural}.

First, the directional coarsening of $\gamma'$ precipitates, the so-called rafting behavior, consists in a large modification of the initial morphology of $\gamma'$ precipitates from cuboidal to elongated or platelet shapes. It has been observed to control the lifetime for the high temperature regime \cite{mughrabi2009microstructural}.  This mechanism has been clearly detailed in the open literature for both experimental and modeling aspects \cite{carry1978apparent, pollock1992creep, cormier2010very, naze2021nickel}. In summary, the authors emphasize from experiments that rafting is the result of both diffusion aspects and the influence of the stress state in the $\gamma$ channel associated with the dislocation \cite{tien1971effect,ignat1993microstructures}. For a negative misfit between $\gamma'$ and $\gamma$ lattice parameters, the dislocations pile up in the channel orthogonal to the applied tensile direction, inducing directional diffusion, and oriented coalescence of $\gamma'$ precipitates \cite{pollock1992creep,chang2018micromechanics}. For such a negative misfit, it has been observed that tensile creep induces rafts orthogonal to the loading direction, N-orientation, whereas compressive creep gives rise to rafts parallel to the loading direction, P-orientation \cite{fredholm1984creep}.

For global optimization of component geometry, especially for hollow high-pressure turbine blades, the thickness of the part could be limited to \SI{0.5}{mm} or less. This can induce a limitation of the lifetime of the component by the so-called thin-wall debit effect \cite{pandey1984environmental}. This effect is due to the fact that the superalloy surface, when exposed to oxidation, is subject to a strong outward flux of Al (and Cr) to form an oxide layer of $Al_2O_3$ (and $Cr_2O_3$). This causes a progressive dissolution of the $\gamma'$ precipitates, resulting in a pure $\gamma$ phase layer up to tens of microns thick. In this way, the thinner the geometry, the greater the reduction in the $\gamma-\gamma'$ surface fraction for a given depletion thickness. This point is reinforced for relatively low stress levels at high temperature: to observe an effect, the time to Al/Cr diffusion should be long enough to influence the microstructure \cite{cassenti2009effect,bensch2013influence}. 

In addition, to reach higher and higher temperatures, thermal barrier coating systems are mandatory for many components \cite{levi2012environmental}. Considering coated Ni-SX, it has been shown that thermal cycling significantly modifies the microstructure evolution of the coating compared to the isothermal conditions often considered in creep loading. Thermo-mechanical fatigue (TMF) is also a key parameter to be analyzed, the lifetime of which is governed by the loading parameters for both bare and coated SX: maximum temperature, high temperature \rev{dwell } time, and stress/strain level \cite{Remy:1993,nutzel2008damage, Sallot:2015, kirka2015influence}. TMF loading could also modify the evolution of the microstructure to a large extent \cite{ai2023thermomechanical}. In addition, coating reduces the thin-wall debit effect for creep loading by limiting the oxidation rate and thus the dissolution of $\gamma'$ \cite{bensch2013influence}. But coating can also drastically affect the lifetime \cite{liu2022coating} by increasing the diffusion flux as a direct function of the respective compositions of the coating and the substrate \cite{zhao2023thickness}.

Regarding modeling aspects, for both rafting and thin-wall debit effect, on the one hand, macroscopic approaches are based on empirical model of creep-damage-microstructure coupling.  Alternatively, the microstructure could be modeled to obtain explicit quantities at the cost of reduced analysis volume. For thin-wall debit effect, most of the open literature proposes  phenomenological models, where the main assumption is based on the loss of a load-bearing section \cite{lv2022stress,mataveli2018thin}: that is, the creep design considers a reduced section, taking into account the thickness of depletion in $\gamma'$. These models may or may not account for the presence of coating \cite{bensch2013influence}.

However, the physics of microstructure evolution is of course more complex than stated in the literature because of the coupling between rafting and depletion, i.e. there is no sharp interface between a depleted zone and a theoretical "perfect" cuboidal $\gamma-\gamma'$ superalloy, since diffusion plays a key role in the $\gamma'$ dissolution process. This can be easily overcome with a macrososcopic model integrating damage analysis (e.g. \cite{cormier2010very}). However, the strong coupling between diffusion, loading parameters and microstructure evolution shows that the interaction prior to damage cannot be neglected. For example, in the case of diffusion coating of SX superalloy, the evolution of coating and IDZ is modified by mechanical parameters for a given thermal cycle \cite{Sallot:2015}.

To address these issues, most advanced models propose to use phase field analysis to account for the coupling between diffusion and mechanical state. However, most people have integrated Fast Fourier Transform (FFT) models for the sake of efficiency (e.g., the work of Steinbach et al. \cite{wang2019combined}), which makes the integration of boundary effects more difficult.
To account for the heterogeneity associated with $\gamma'$ depletion, a very interesting proposal has recently been made, which first considers different microstructure and phase field analyses \cite{yu2020thickness}. In this way, the authors demonstrate the influence of the initial $\gamma-\gamma'$ morphology on the stress state, but are limited to a somewhat "static" view of a dynamic process.  

First, a brief introduction is given to the experimental evidence of complex coupling between diffusion, phase transformation and mechanical coupling for both bare and coated SX superalloys. Then, the aim of the study is to propose an original phase field model, implemented in an FE code, \rev{accounting for } the complex loading in the presence of diffusion flux, in order to describe the $\gamma-\gamma'$ microstructure evolution considering only the applied loading and flux boundary conditions in the vicinity of coated or free surfaces.

\section{Experimental results in TMF}

\subsection{Materials, specimens and loading}
For the experimental aspects, the analysis is carried out on tests from the work of Pierre Sallot, detailed in \cite{Sallot:2015}. The substrate is a first generation SX Ni-based superalloy (AM1), the coating is (Ni,Pt)Al obtained by electrodeposition of Pt and vapor phase diffusion of Al (details are given in \cite{Mevrel:1987,Sallot:2015} for the so-called aluminum phase vapor deposition Safran (APVS) process). To control the TMF \rev{test}, the samples consist of hollow specimens where the coated surface is the outer surface and the inner surface is only machined \cite{Sallot:2015}. 
\rev{Heating for the TMF test is achieved using a four-lamp furnace equipped with forced air cooling. To control the temperature, for both heating and cooling, by power delivered to the lamps and air flow respectively, a thermocouple is fixed within the gauge length, controlling the temperature by a closed loop \cite{Koster:1994}. The temperature gradients, both axial and radial, are less than 5 K \cite{Maurel:2010a}. Since the TMF test is stress controlled, only cell force is used in this case}. 

The applied \rev{temperature } consists of thermal cycling from 100 to \SI{1100}{\celsius} with a dwell time of 5 min at \SI{1100}{\celsius}, cooling and heating being linear for 5 min each. TMF cycles were applied using the same thermal cycles with stress dwell in tension or compression, called in-phase (IP) and out-of-phase (OP) tests, respectively. We focus in this paper on a condition consisting of IP-TMF for \SI{30}{MPa} applied during the dwell at the maximum temperature of the test.



\subsection{Microstructure evolution in presence of Al-rich coating and for bare surface}
\begin{figure}[!h]
	\centering
		\includegraphics[width=\columnwidth]{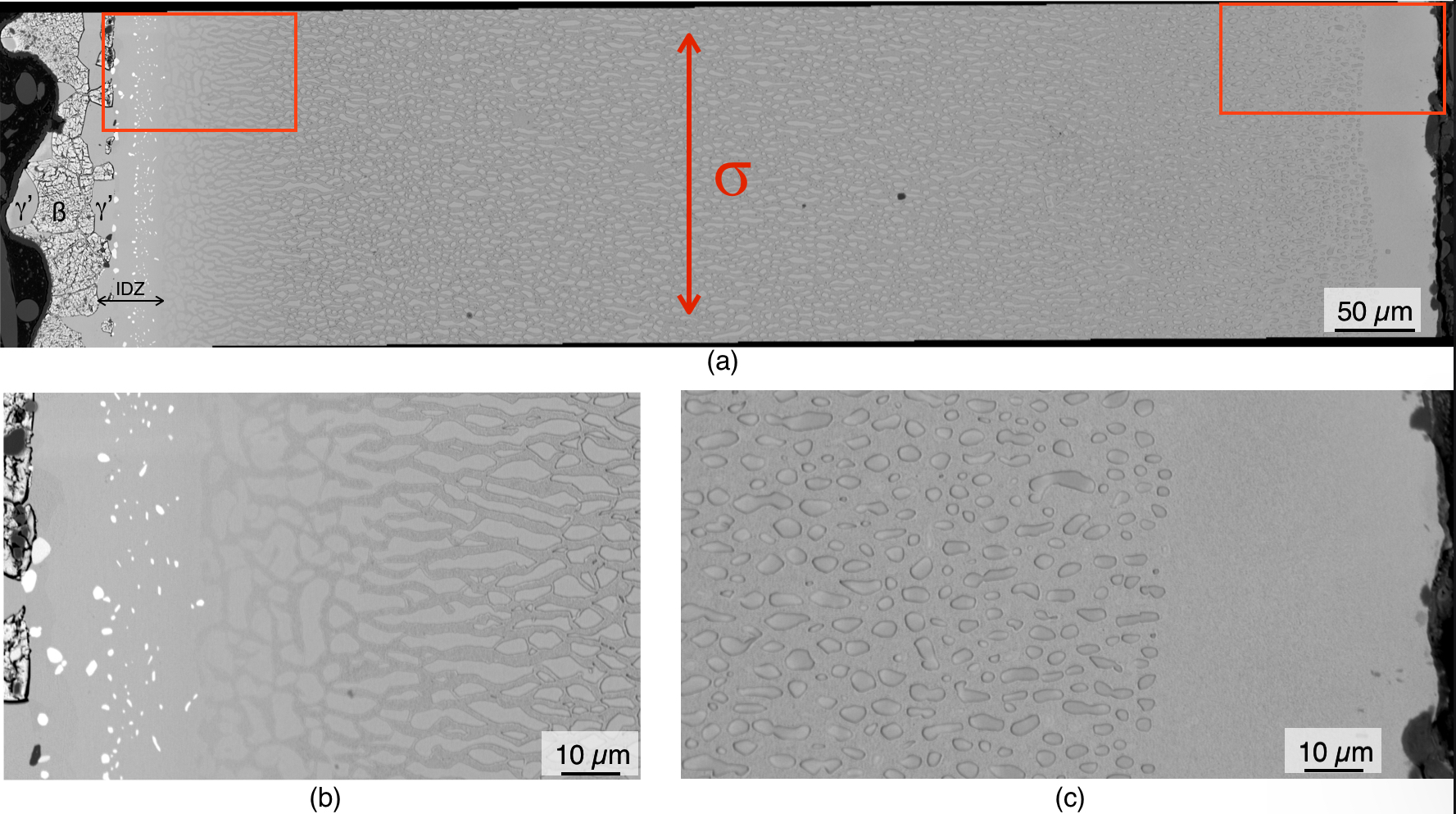}
	\caption {Cross-section parallel to the loading direction  observed in BSE mode, for IP-TMF test (30 MPa at \SI{1100}{\celsius} for 5 min) corresponding to 2000 cycles (a) global view through the thickness of the specimen (b) detail close to the IDZ (c) detail close to the bare surface} 
 \label{fig:expIP302}
\end {figure}

As an example, \rev{we detail the } the IP-TMF test considered, \rev{after 2000 cycles (that are equivalent to 166 h spent at \SI{1100}{\celsius})}: induced rumpling (surface roughness increase) together with the $\beta$ to $\gamma'$ phase transformation in the outer part of the coating are obvious, Figure \ref{fig:expIP302}(a).  Looking at the coating/substrate interface,  the rafting is strongly modified by the presence of the coating: close to the IDZ the rafts are more or less parallel to the stress direction (here vertical), P-type rafting, whereas far from the IDZ the rafts are orthogonal to the stress direction, N-type rafting, see Figure \ref{fig:expIP302}(b). \rev{It is worth noting that this area also corresponds to the area of contrast inversion induced by the Pt content, which is known to modify misfit and subsequently rafting behavior \cite{reed2008}}.


Of great interest is the evolution of the microstructure near the uncoated surface of the sample, which is prone to oxidation, Figure \ref{fig:expIP302}(c).  Since the oxide is mainly alumina, it can be assumed that an outward flux of Al is promoted at this location (inner surface of the hollow specimen). The observation of the microstructure is obvious: the depletion of Al leads to the ovalization of the $\gamma'$ precipitates, the increase of the $\gamma$ channel width and the final dissolution of the $\gamma'$ precipitates.  In addition, rafting occurs in competition with the dissolution of $\gamma'$ precipitates. Compared to the mid-thickness of the substrate, Figure \ref{fig:expIP302}(a), where N-type rafts are observed, the influence of the outward flux of Al is evident, Figure \ref{fig:expIP302}(c).

This point raises the question of how diffusion flux and rafting, which are a priori driven by lattice strain mismatch and creep state, are coupled.

\vspace{-2em}
\section{Fully coupled model based on phase-field}
\subsection{Basis of the model}

To model the evolution of the $\gamma-\gamma'$ morphology under TMF loading conditions, a phase field model developed by Cottura et al. \cite{cottura2012phase} is applied, so as to describe strong diffusion-mechanical coupling.

First, the problem should define the evolution of the concentration of species in the system, which is here limited to the Al content and will be denoted by $c$, i.e. the model will describe the binary Ni-Al but for only one diffusion coefficient \cite{cottura2012phase}. Second, the phase field will describe either $\gamma$ or $\gamma'$ phases, allowing for four different variants of $\gamma'$ phases. Three non-conservative associated order parameters are used to describe the four variants of the $\gamma'$ phase and are denoted by $\phi_i$ (i=1 to 3), see table \ref{tab:variants} for details of triplets and variants. The main modification of the model is the use of the finite element formulation instead of the FFT. This allows the model to easily account for boundary effect, which is mandatory to consider the above cases of oxidation and/or coating. 

\begin{table}[h]
   	\centering
    \caption{Order parameters triplets and associated $\gamma$ or $\gamma'$ phases, and variants}
			\begin{tabular}{ ccc cc }
				\hline
				Phase & Variant & $\phi_1$ & $\phi_2$ & $\phi_3$  \\
				\hline
				$\gamma$     & - & 0 & 0 & 0  \\
				\hline
				\multirow{4}{*}{$\gamma^{'}$} & 1 & 1 & 1 & 1  \\
				& 2 & -1 & -1 & 1  \\
				& 3 & 1 & -1 & -1  \\
				& 4 & -1 & 1 & -1  \\
				\hline
			\end{tabular}
         \label{tab:variants}
  \end{table}

The proposed model is based on anisotropic elasticity for both the $\gamma$ and $\gamma'$ phases. The $\gamma'$ phase is described by linear elasticity, while the $\gamma$ phase is modeled by nonlinear kinematic hardening, which will be detailed in the sequel. Thus, both elastic strain tensor and internal variables are introduced.

 On this basis, the global model is described by the following set of variables:
 \begin{itemize}
	\item Al concentration $c$,
	\item Order parameters $\phi_i$ and their gradients $\nab \phi_i$ ($i=1,2$ and $3$),
	\item Elastic strain tensor $\ten \varepsilon^e$,
	\item Internal variables $V_k$ associated with strain hardening ($V_k = \left( \ten \alpha, p \right)$ for kinematic and isotropic strain hardening, respectively).
\end{itemize}

Finally, the phase field model is described by its mesoscopic energy, given by
\begin{eqnarray*}
F=F(c,\phi_i,\nab\phi_i,\ten \varepsilon^e,\ten \alpha,p)  =  \int_V f(c,\phi_i, \nab\phi_i,\ten \varepsilon^e,\ten \alpha,p) dv\\
=  \int_V \left[ f_{chem}(c,\phi_i) + f_{mech}(c,\phi_i,\ten \varepsilon^e, \ten \alpha,p) + \dfrac{\beta}{2} \sum_{i=1}^{3} \mid \nab \phi_i \mid^2 \right] dv
\end{eqnarray*}
where the chemical energy $F_{chem}$ depends on the Al concentration $c$ and the order parameter $\phi_i$, the mechanical energy $F_{mech}$ depends on $c$, $\phi_i$ but also on the elasticity and the hardening behavior described by $\ten{\varepsilon}^{e}$, $\ten{\alpha}$ and $p$. From this free energy, the state laws for concentration and order parameters are derived.

It is also worth noting that strain compatibility allows to consider an  eigenstrain tensor associated with the phase transformation from $\gamma$ to $\gamma$', namely $\ten{\varepsilon}^*$:
\begin{equation}
	\ten{\varepsilon}=\ten{\varepsilon}^{e} +\ten{\varepsilon}^{vp} +\ten{\varepsilon}^{*}(c)
\end{equation}
The phase transformation tensor is described as:
\begin{equation}
	\ten{\varepsilon}^*(c)=\varepsilon^{T}(c-c^{eq}_{\gamma}) \ten{1}
\end{equation}
where $\varepsilon^{T}=\frac{\delta}{c^{eq}_{\gamma'}-c^{eq}_{\gamma}}$ is the strain induced by the lattice mismatch $\delta=2 \frac{a_{\gamma'}-a_{\gamma}}{a_{\gamma'}+a_{\gamma}}$, where $c^{eq}_{\gamma}$ and $c^{eq}_{\gamma'}$ \rev{are the Al concentrations in the $\gamma$ and $\gamma'$ phases, respectively}.

\subsection{Chemical Free Energy}
Following \cite{cottura2012phase}, the chemical free energy is described as:
\ 
\begin{eqnarray}
F_{chem}(c,\phi_i)  =  \int_V f_{chem}(c,\phi_i) dV \nonumber\\
 =  \int_V \Delta f \Big[ \frac{1}{2} \left(c-c_{\gamma} \right)^2 + \frac{B}{6} \left(c_2-c \right) \sum_{i=1}^3 \phi_i^2 - \frac{C}{3} \phi_1 \phi_2 \phi_3 + \frac{D}{12} \sum_{i=1}^3 \phi_i^4 \Big] dv 
\label{F_chim}
\end{eqnarray}

with:
\begin{eqnarray}
\left\{
\begin{aligned}
B &= 2 \left( c_{\gamma\prime} - c_{\gamma} \right)
\\
C &= 6 \left( c_{\gamma\prime} - c_{\gamma} \right) \left( c_2 - c_{\gamma} \right)
\\
D &= 2 \left( c_{\gamma\prime} - c_{\gamma} \right)\left( c_{\gamma\prime} + 2 c_2 - 3 c_{\gamma} \right)
\end{aligned}\right. 
\end{eqnarray}

\begin{figure}[!h]
	\centering
	\includegraphics[width=0.55\columnwidth]{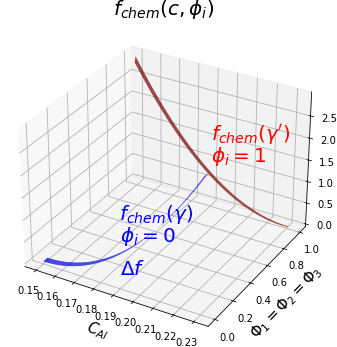}
	\caption{Free energy value for $\gamma$ and $\gamma'$ state highlighting the case of the variant of $\gamma'$ corresponds to $\phi_1=\phi_2=\phi_3=1$}
	\label{fig:mesh}
\end{figure}

The free energy $f_{\gamma}$ (or $f_{\gamma'}$) of the $\gamma$ (or $\gamma$') phase is given as a function of Al concentration by equation (\ref{F_chim}), for which the order parameters are set to their equilibrium values for the $\gamma$ (or $\gamma$') phase, Figure \ref{fig:mesh}.

\subsection{Mechanical Free Energy}
\medskip

The second contribution to the free energy density is due to mechanical effects. Assuming that elastic behaviour and hardening are uncoupled, the mechanical free energy $F_{mech}$ is decomposed into a coherent elastic energy density $F_{el}(c,\ten \varepsilon^e)$  and a viscoplastic part  $F_{vp}(\phi_i, \ten \alpha,p)$ as:

\begin{eqnarray}
F_{mech}(c,\phi_i,\ten \varepsilon^e, \ten \alpha,p) = F_{e}(c,\ten \varepsilon^e) + F_{vp}(\phi_i, \ten \alpha,p)
\end{eqnarray}

The elastic part $F_{e}$ is given by the relation (\ref{Fel}) :
\begin{equation}
F_{e}(c,\ten \varepsilon^e) = \frac{1}{2}\int_V \ten \varepsilon^e:\ordq \lambda(c):\ten \varepsilon^e dV
\label{Fel}
\end{equation}
The elasticity tensor $\ordq \lambda$ is interpolated by the function $h(c)$ :
\begin{equation}
\ordq \lambda = h(c)\ordq \lambda^{\gamma'} + [ 1-h(c)]\ordq \lambda^{\gamma}
\end{equation}

where the interpolation function $h$ is described by  the following expression:
\begin{equation}
h(c) = \frac{c - c_{\gamma}}{c_{\gamma'} - c_{\gamma}}
\end{equation}

The tensors of elasticity of the cubic materials used for the two  $\gamma$ and $\gamma$' phases have the following form:

\begin{equation}
	\lambda_{ijkl} = \begin{bmatrix}
	C_{11} & C_{12} & C_{12} &   0    &   0    &   0    \\
	& C_{11} & C_{12} &   0    &   0    &   0    \\
	&        & C_{11} &   0    &   0    &   0    \\
	&        &        & C_{44} &   0    &   0    \\
	&  Sym   &        &        & C_{44} &   0    \\
	&        &        &        &        & C_{44} \\
	\end{bmatrix}
\end{equation}

where $Sym$ indicates the symmetry used for the elastic tensor.


The viscoplastic part $F_{vp}$ is interpolated at the interface between the $\gamma$ and $\gamma'$ phases by the interpolation function $h_{mech}$. It is described by the following equations:

\begin{eqnarray}
F_{vp}(\phi_i,\ten \alpha,p) = h_{mech}(\phi_i) F_{vp}(\ten \alpha,p)^{(\gamma')} \nonumber \\+ \left( 1-h_{mech}(\phi_i) \right) F_{vp}(\ten \alpha,p)^{(\gamma)}
\end{eqnarray}

\begin{equation}
\textrm{with }F_{vp}(\underset{\sim}{\alpha},p)^{(k)} = \int_V \left[ \frac{1}{3} C^{(k)} \underset{\sim}{\alpha}:\underset{\sim}{\alpha} + \frac{1}{2} H^{(k)} p^2 \right] dV 
\end{equation}
where $k=(\gamma,\gamma')$, and
\begin{equation}
h_{mech}(\phi_i) = \frac{1}{2} + \frac{1}{2} \tanh \left[ \theta \left( \frac{1}{3} \sum_{i=1}^3 \phi_i^2 - \frac{1}{2} \right) \right] 
\end{equation}
where $\theta=100$.

A simple quadratic form of all internal state variables $p$ and $\underset{\sim}{\alpha}$ is assumed for the hardening contribution.

Once the free energy density are defined, the Cauchy stress tensor and the thermodynamic driving forces $\ten X$ and $R$ associated respectively with the internal kinematic and isotropic hardening variables $\ten \alpha$ and $p$, are

\begin{gather*}
\ten \sigma = \dfrac{\partial f_{e}}{\partial \ten \varepsilon^e} = \ordq \lambda(c): \ten \varepsilon^e
\\
\ten X = \dfrac{\partial f_{vp}}{\partial \ten \alpha} = h_{mech}(\phi_i) \frac{2}{3} C^{(\gamma')} \ten \alpha + \left( 1 - h_{mech}(\phi_i) \right) \frac{2}{3} C^{(\gamma)} \ten \alpha
\\
R = \dfrac{\partial f_{vp}}{\partial p} = h_{mech}(\phi_i) \frac{2}{3} H^{(\gamma')} p + \left( 1 - h_{mech}(\phi_i) \right) \frac{2}{3} H^{(\gamma)} p
\end{gather*}

In order to describe  the irreversible part of the mechanical behaviour and to define the complementary laws related to the dissipative process, a convex dissipation potential is postulated, which depends on stress tensor  $\ten \sigma$  and the thermodynamic forces $\ten X$ and $R$.



The complementary evolution laws of the internal variables are derived from the dissipation potential as:

\begin{gather*}
\dot{p}=\left\langle \frac{J_2(\ten \sigma - \ten X)-R_0-R}{K} \right\rangle^N=\sqrt{\frac{2}{3}\dot{\ten \varepsilon}^p:\dot{\ten \varepsilon}^p}
\\
\dot{\ten \alpha}=\dot{\ten \varepsilon}^p-D\ten \alpha \dot{p}
\\
\dot{\ten \varepsilon}^p=\frac{3}{2} \dot{p} \frac{\ten \sigma^{'} - \ten X^{'}}{J_2 \left(\ten \sigma^{'} - \ten X^{'}\right)}
\end{gather*}

In the present work, a von Mises criterion is adopted:

\begin{equation}
J_2(\ten\sigma)=\sqrt{\frac{3}{2}\dot{p}\left(\ten \sigma^{'}:\ten \sigma^{'}\right)} 
\end{equation}
$\textrm{with } \ten \sigma^{'} = \ten \sigma - \frac{1}{3} tr(\ten \sigma):\ten 1 \textrm{ deviatoric term of  } \ten \sigma$, 
where $R_0$ is the initial yield strength. This is a simplification compared to crystal plasticity modeling of single crystal superalloys. This simplifying assumption was also adopted in \cite{cottura2012phase}. In contrast, crystal plasticity was used in \cite{cottura2016coupling}.

\subsection{Evolution equations}

The evolution of the conservative concentration field is described by the diffusion equation:
\begin{eqnarray}
\displaystyle  \displaystyle  \dot{c} = - \nab . \vect J = - \nab . \left(-L \nab \mu \right) \nonumber \\= L \Delta \left( \frac{\partial f_{chem}}{\partial c}(c,\phi_i)\ + \frac{\partial f_{mech}}{\partial c}(c,\phi_i,\ten \varepsilon^{e},\ten \alpha,p) \right)
\end{eqnarray}
where $\vect J$ is the diffusion flux,  $L$ is the Onsager coefficient, which is related to the diffusion coefficient $D$ by:
\begin{equation}
L = D \left(\frac{\partial^2 f_{chem}}{\partial c^2} \right)^{-1} \quad \textrm{ and } D = D_0 \exp{ \left[ \frac{-\Delta U}{k_{\beta} T}\right]}
\end{equation}

For a given temperature, these quantities are constant. The evolution of the non-conservative field of order parameters is modeled by the Allen-Cahn equation, with M a constant kinetic coefficient related to interfacial mobility:

\begin{equation}
\displaystyle  \dot{\phi_i} = -M \left( \frac{\partial f_{chem}}{\partial c}(c,\phi_i) +  \frac{\partial f_{mech}}{\partial c}(c,\phi_i,\ten \varepsilon^{e},\ten \alpha,p) - \beta \Delta \phi_i\right) 
\end{equation}
with $(i=1,2,3)$.

During diffusive phase transformations, as in the case of Ni-based superalloys, the elastic energy equilibrates much faster than the characteristic diffusion time. Therefore, when studying microstructural evolution, it is possible to assume that mechanical equilibrium is always maintained. In the case of microstructural evolution under applied load, we have :
\begin{equation}
{\rm div}(\ten \sigma)+ \vect f = div \left(\dfrac{\partial f_{mech}}{\partial \ten \varepsilon^{e}}(c,\phi_i,\ten \varepsilon^{e},\ten \alpha,p) \right)+ \vect f =0 
\end{equation}
in the volume V, and
\begin{equation}
 \vect t = \ten \sigma.\vect n \textrm{ at the surface $\partial$V}
\end{equation}
volume forces $\vect f$ are not considered in the following.
\vspace{-1em}
\section{Application to a diffusion problem with or without applied stress}
\subsection{Model and boundary conditions}
\begin{figure}[!h]
		\centering
	   \includegraphics[width=\columnwidth,angle=0]{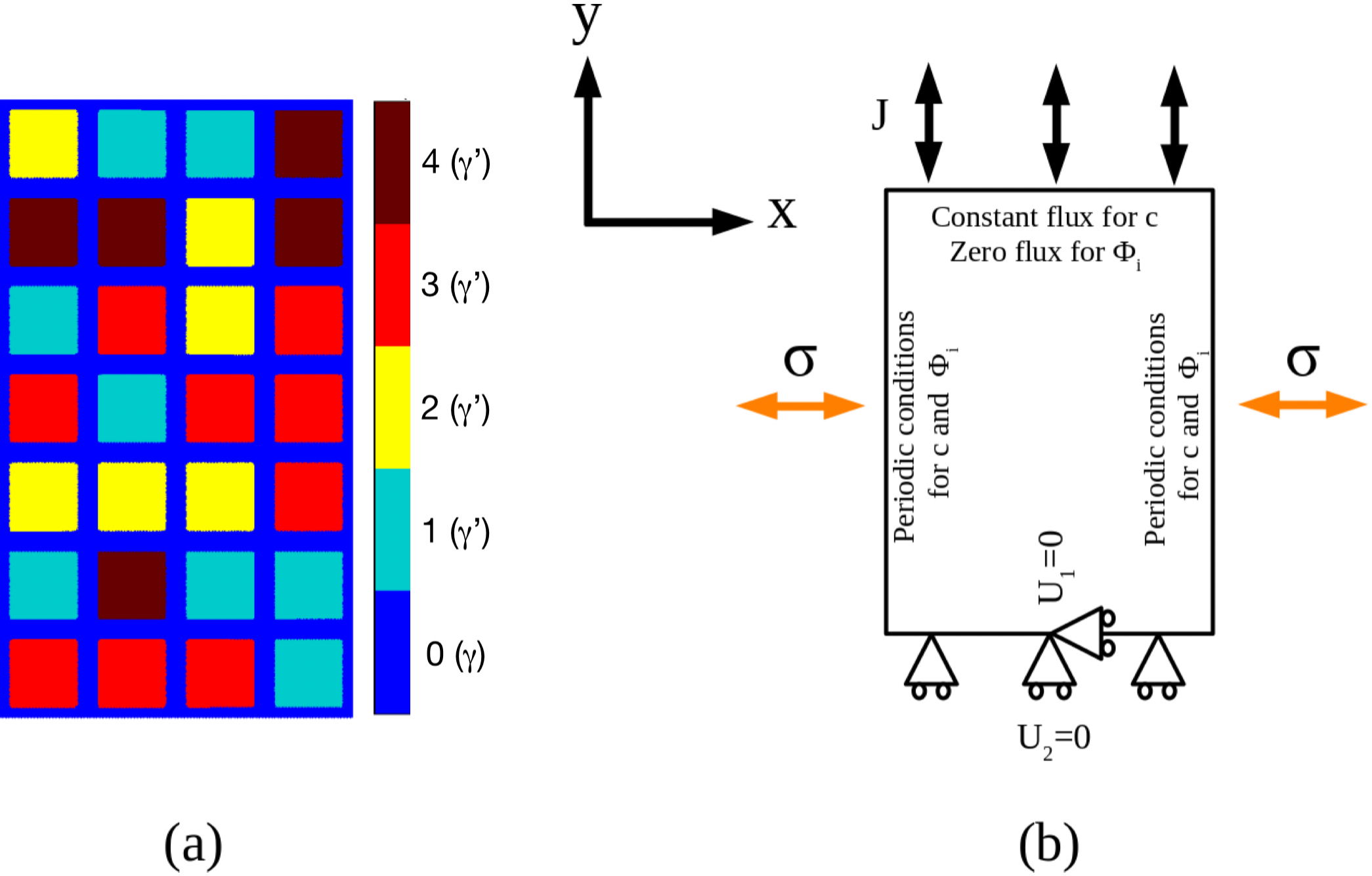}  
	\caption{(a) initial $\gamma-\gamma'$ microstructure, each variant being associated to a color (see colorbar and Table \ref{tab:variants}) (b) mechanical and chemical boundary conditions for the coupled problem}
	\label{fig:sim}
\end{figure}

We apply the above phase field model to a simple geometry representative of a $\gamma-\gamma'$ microstructure: The modeled microstructure consists of a 4x7 $\gamma'$ precipitates with an initial square shape, Figure \ref{fig:sim}(a). Each variant is described by a specific color, with the blue color corresponding to the $\gamma$ matrix.  The edge of the $\gamma'$ precipitates is set to \SI{420}{nm}, for a channel width of \SI{80}{nm}. The total size of the model is 2x3.5 $\mu m^2$. Using a 2.5D plane strain assumption, this model induces 400,000 degrees of freedom.
Experimental results show that there is a strong interaction between the $\gamma-\gamma'$ microstructure and the diffusion fluxes. Based on the above model, we consider only the Al flux in the following. This Al flux could be either inward or outward to model the case of an Al-rich coating (similar to the example shown in Figure \ref{fig:expIP302}(b)) and a free surface exposed to oxidation (Figure \ref{fig:expIP302}(c)), respectively. For simplicity, the Al flux is considered constant in this analysis.
The boundary conditions are as shown in figure \ref{fig:sim}(b):
\begin{itemize}
    \item periodic conditions at edges normal to $x$ \rev{for both Al flux and displacement},  together with applied force conditions (see orange arrows in figure \ref{fig:sim}(b));
    \vspace{-.5em}
    \item a zero net flux at the edge normal to $y$ (here at the bottom of figure \ref{fig:sim}(a)), \rev{with a zero displacement in direction $y$ for the same boundary};
    \vspace{-.5em}
    \item an outward flux of Al at the opposite edge normal to $y$ (see "J" black arrows in figure \ref{fig:sim}(b)), here set to a constant value $J_0$=5.$10^{-6}$ mm$^{-2}$h$^{-1}$.
\end{itemize}
\vspace{-.5em}
Prior to applying the TMF loading conditions, an isothermal dwell of 1 hour at \SI{950}{\celsius} is applied to mimic the high temperature heat treatment typical of this type of SX superalloy.

The proposed analysis is based on the isothermal evolution of a typical $\gamma-\gamma'$ superalloy. This assumption is made for simplicity, particularly in the absence of data to mimic the role of low-temperature plasticity on diffusion. The proposed analysis will then consider only the diffusion-mechanical coupling for different fluxes of Al and different creep loads. To gain understanding in the context of the TMF loading described above, two different scenarios of diffusion with or without creep loading will be detailed: the outward flux of Al, modeling the oxidation condition on the free (inner) surface of the specimen, and the inward flux of Al, modeling the Al-rich coating condition.

Considering the mechanical behavior of the $\gamma-\gamma'$ superalloy, the values correspond to the nickel-based single crystal superalloy AM1 according to \cite{cottura2012phase}, see table \ref{tab:parambehavior}.

\begin{table*}[!h]
    \centering
    \caption{Coefficients for mechanical behavior, H=0 for both phases, assuming constant isotropic hardening and non-linear kinematic hardening}
	\begin{tabular}{ccc ccc ccc }
		\hline
		Phases & $C_{11}$(GPa) & $C_{12}$(GPa) & $C_{44}$(GPa) & C(GPa) & D & n & $K(MPa.s^{1/n})$ & $R_0(MPa)$  \\
		\hline
		$\gamma$     & 197 & 144 & 90 & 150 & 1900 & 5 & 150 & 86      \\ 
		$\gamma^{'}$ & 193 & 131 & 97 & 150 & 1900 & 5 & 150 & $10^5$  \\ 
		\hline
	\end{tabular}
 \label{tab:parambehavior}
\end{table*}
Setting H=0 for both phases assumes constant isotropic hardening and nonlinear kinematic hardening. The constitutive equations are the same for both phases, but a very high yield stress $R_0$ is set for $\gamma'$ so that its response remains purely elastic in the simulations.



\subsection{Oxidation analyzed through outward flux of Al}
\begin{figure}[h!]
		\centering
	   \includegraphics[width=0.85\columnwidth,angle=0]{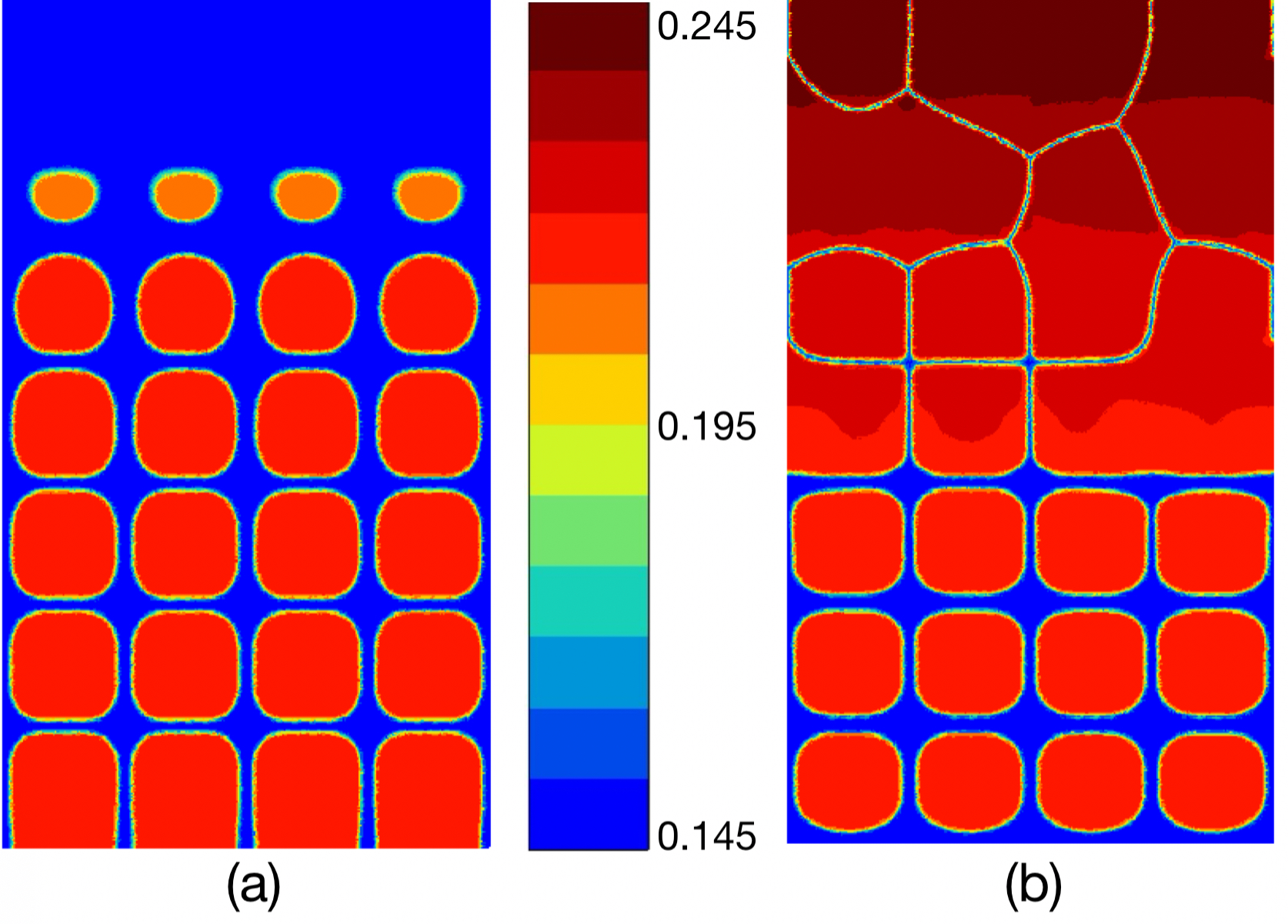}
	\caption{Al concentration  for $|J|=J_0$ without applied stress (a) for outward Al flux and (b)  for inward Al flux (for \SI{9.8}{h} of diffusion)}
	\label{fig:Jnosig}
\end{figure}

To model a case similar to pure oxidation of a free surface, a constant outward flux of Al is prescribed, Figure \ref{fig:Jnosig}(a). Here, the model is simplified for reasons of computational time by using only elasticity for coupling (i.e. viscoplasticity is not considered yet and  a very high yield stress $R_0$ is set also for $\gamma$ phase).

Similar to experimental results, $\gamma'$ precipitates dissolve progressively, Figure \ref{fig:Jnosig}(b): the morphology of the $\gamma'$ phases becomes more or less circular, and $\gamma'$ precipitates dissolve into the $\gamma$ matrix, with an obvious dissolution gradient. Depending on the prescribed $\gamma'$ variant, coalescence of $\gamma'$ precipitates before dissolution is not observed, Figure \ref{fig:Jnosig}(b). These observations are consistent with the experiment when the side of the sample directly exposed to oxidation is considered, Figure \ref{fig:expIP302}(c). It is worth noting that the gradients of Al concentration seem to be confined to the $\gamma-\gamma'$ interfaces.


\subsection{Diffusion coating analyzed through inward flux of Al}
To model a case similar to a rich Al coating, a constant inward flux of Al is prescribed in Figure \ref{fig:sim}(b), \rev{for J oriented toward negative $y$ values}. Again, only elasticity is considered for this case.

Here a coalescence of $\gamma'$ is obviously obtained, Figure \ref{fig:Jnosig}(b). This observation is consistent with the experiment considering the vicinity of the IDZ, Figure \ref{fig:expIP302}(b). For the same diffusion time (here \SI{9.8}{h}), this inward flux seems to have a more drastic effect on the microstructure than the outward flux of Al, compare figures \ref{fig:Jnosig}(a) and (b).  Moreover, within the dense $\gamma'$ region (upper part of figure \ref{fig:Jnosig}(b)), the gradients of Al are evident, with local interactions with the variant of $\gamma'$ and a very small channel of $\gamma$ remaining between the almost coalesced $\gamma'$ phases.


\subsection{Creep and outward Al flux}
\begin{figure}[ht]
		\centering
	   \includegraphics[width=0.8\columnwidth,angle=0]{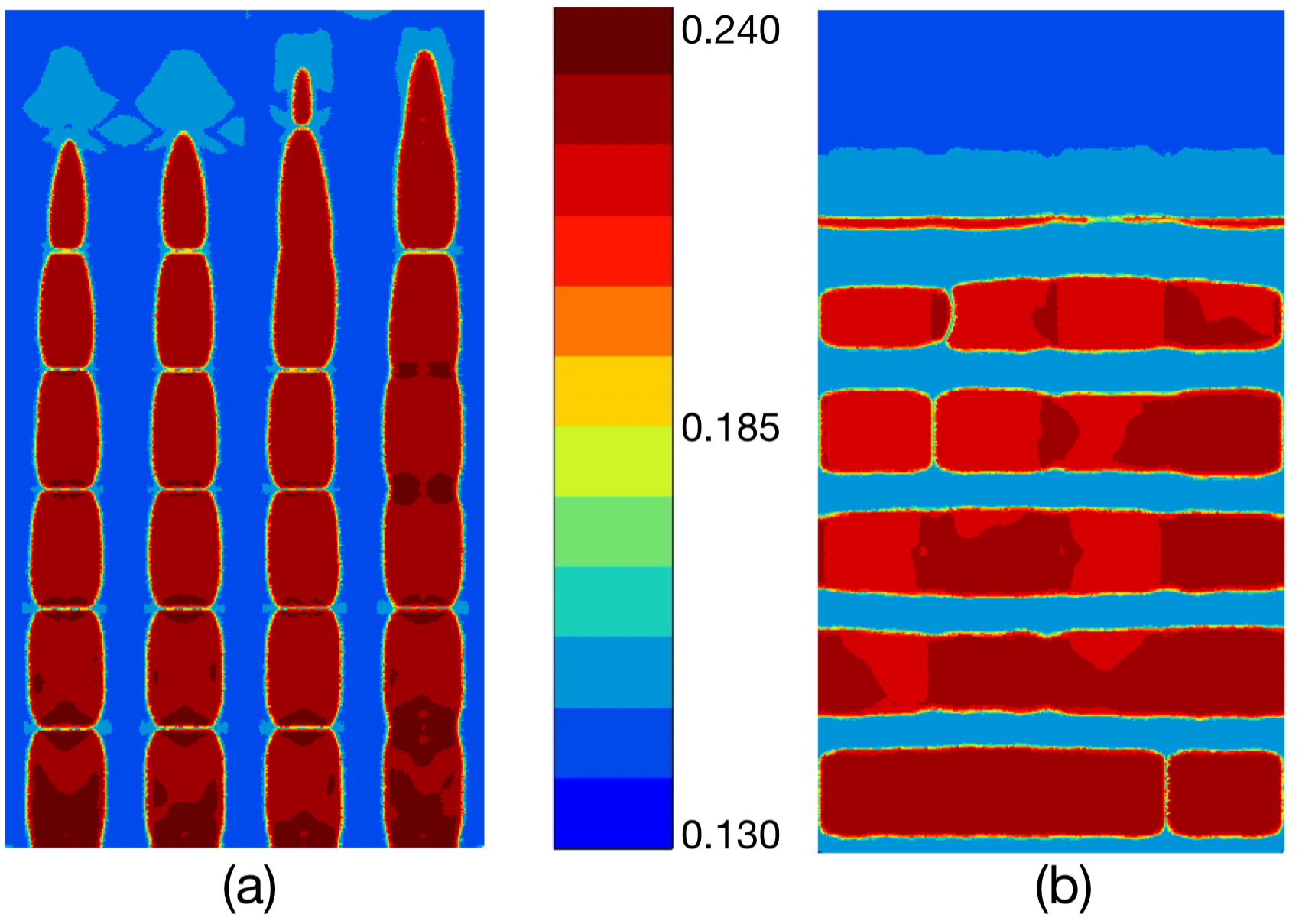}
	\caption{Al concentration for outward  Al flux $J=J_0$,  (a) for tensile creep and (b) for compressive creep  (for \SI{9.8}{h} of diffusion)}
	\label{fig:Joutwsig}
\end{figure}To get a clearer view of the chemical/mechanical coupling at work in $\gamma-\gamma'$ superalloys, the addition of stress at high temperature is required. The focus is on the analysis of the outward flux of Al to mimic the oxidation of a superalloy. In this case, we consider the full model, taking into account the elasto-viscoplasticity in the $\gamma$ phase.

As a first analysis, we describe a creep stress value set to \SI{240}{MPa} and an Al flux $J=J_0$ for both tension and compression, Figure \ref{fig:sim}(b), \rev{for J oriented toward positive $y$ values}.

\begin{figure}[ht] 		\centering
	   \includegraphics[width=1\columnwidth]{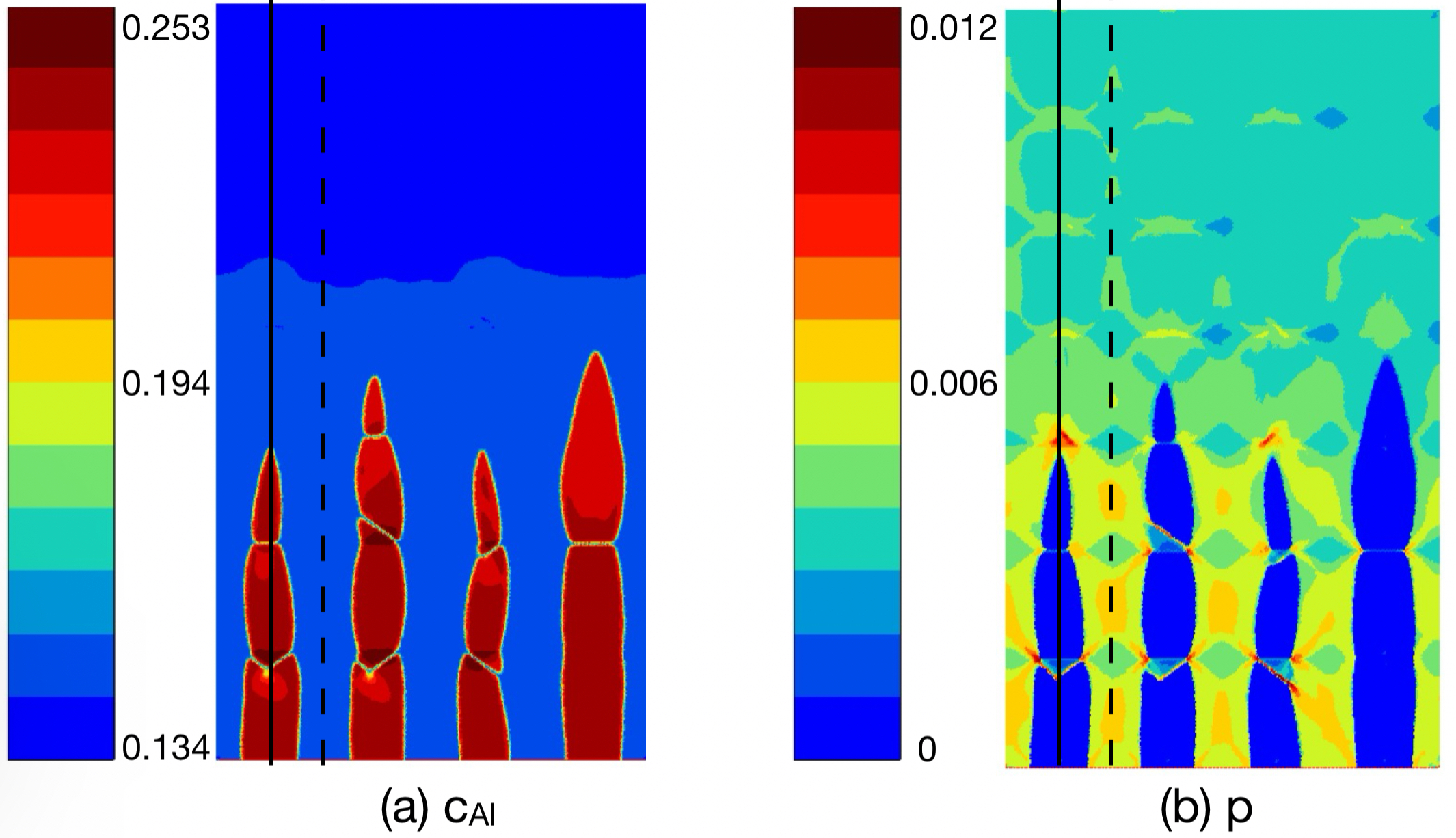}
	\caption{Outward Al flux and  tensile creep (a) Al concentration and (b) cumulated plastic strain (for \SI{13}{h} of diffusion)}
	\label{fig:JoutwsigCetP}
\end{figure}

For a tensile creep stress, N-type rafts are obtained along with the dissolution of $\gamma'$, Figure \ref{fig:Joutwsig}(a). It is noteworthy that in this case a strong gradient of Al concentration is obtained both in the $\gamma$ matrix and within the $\gamma'$ precipitates. Moreover, the morphology of the precipitates is very different compared to the case where only diffusion has been considered, compare figures \ref{fig:Jnosig}(a) with \ref{fig:Joutwsig}(a): the application of a stress at high temperature induces a more elongated structure of the precipitates before their coalescence, a strong boundary effect is observed for precipitates before their dissolution (precipitates closer to the boundary normal to $y$ at the top of the model).  The variants of $\gamma'$ precipitates affect their dissolution in such a way that there is a scattering in the four rows of precipitates modeled here. Also, the rafts appear to "oscillate" slightly, in the sense that they are not perfectly straight-sided and display a wavy microstructure. This last point is consistent with most experimental observations of rafting and is more pronounced at longer exposure times, Figure \ref{fig:JoutwsigCetP}(a).


For compressive creep, P-type rafts are obtained along with the dissolution of $\gamma'$, Figure \ref{fig:Joutwsig}(b). Compressive stress leads to a more pronounced dissolution of $\gamma'$ than tensile stress for the same diffusion creep time. In addition, the morphology of the precipitates is more "rounded" in compression than in tension, although wavy rafts are also observed in compression.

These large changes in the morphology of the precipitates depending on the sign of the applied stress require further analysis of the coupling and will be discussed in the sequel.

\vspace{-2.5em}
\section{Discussion}
\subsection{Localization Aspects of Strain and Diffusion}
Together with the Al concentration, the proposed model allows the analysis of mechanical quantities. For longer exposure times to diffusion and creep, both the composition in Al and the associated cumulative plastic strain field are plotted respectively in Figures \ref{fig:JoutwsigCetP}(a) and (b). Considering the composition, this snapshot again shows gradients of rather low amplitude (long range of Al concentration variation) in the $\gamma$ matrix and higher amplitude (shorter range of Al concentration variation) in the $\gamma'$ precipitates, Figure \ref{fig:JoutwsigCetP}(a). Since we are considering cumulated plasticity here, the map corresponds to a transformation history, Figure \ref{fig:JoutwsigCetP}(b): previous locations of $\gamma'$ precipitates are evidenced by the strain localization in the matrix around these $\gamma'$ precipitates before their dissolution. In addition, the narrow morphology of the precipitates induces local strain maxima, e.g. for the first and third rows of precipitates from the left.

\begin{figure}[ht]
    \centering
    \begin{subfigure}[b]{.8\columnwidth}
        \centering
        \includegraphics[width=\columnwidth]{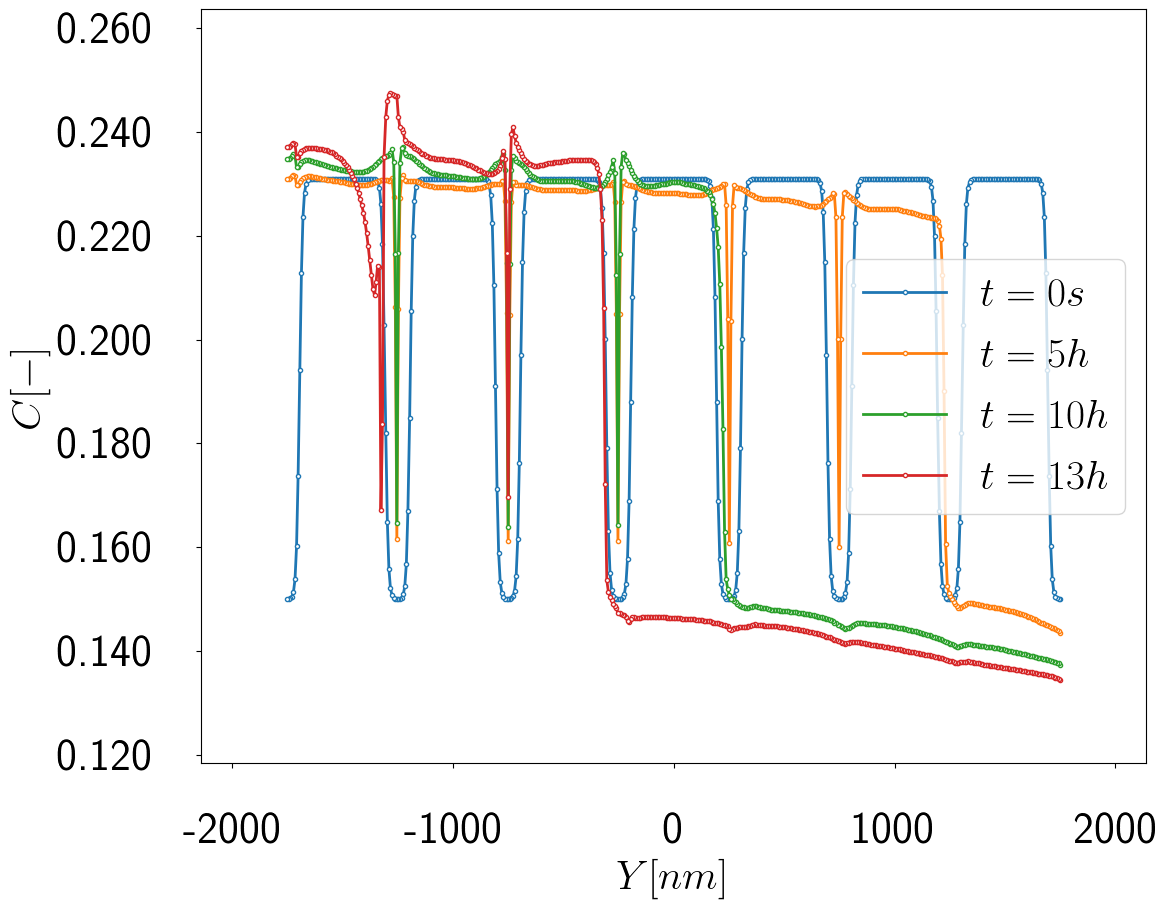}
        \caption{Al concentration \label{fig:Alpr}}
    \end{subfigure}
    \begin{subfigure}[b]{.8\columnwidth}
        \centering
        \includegraphics[width=\columnwidth]{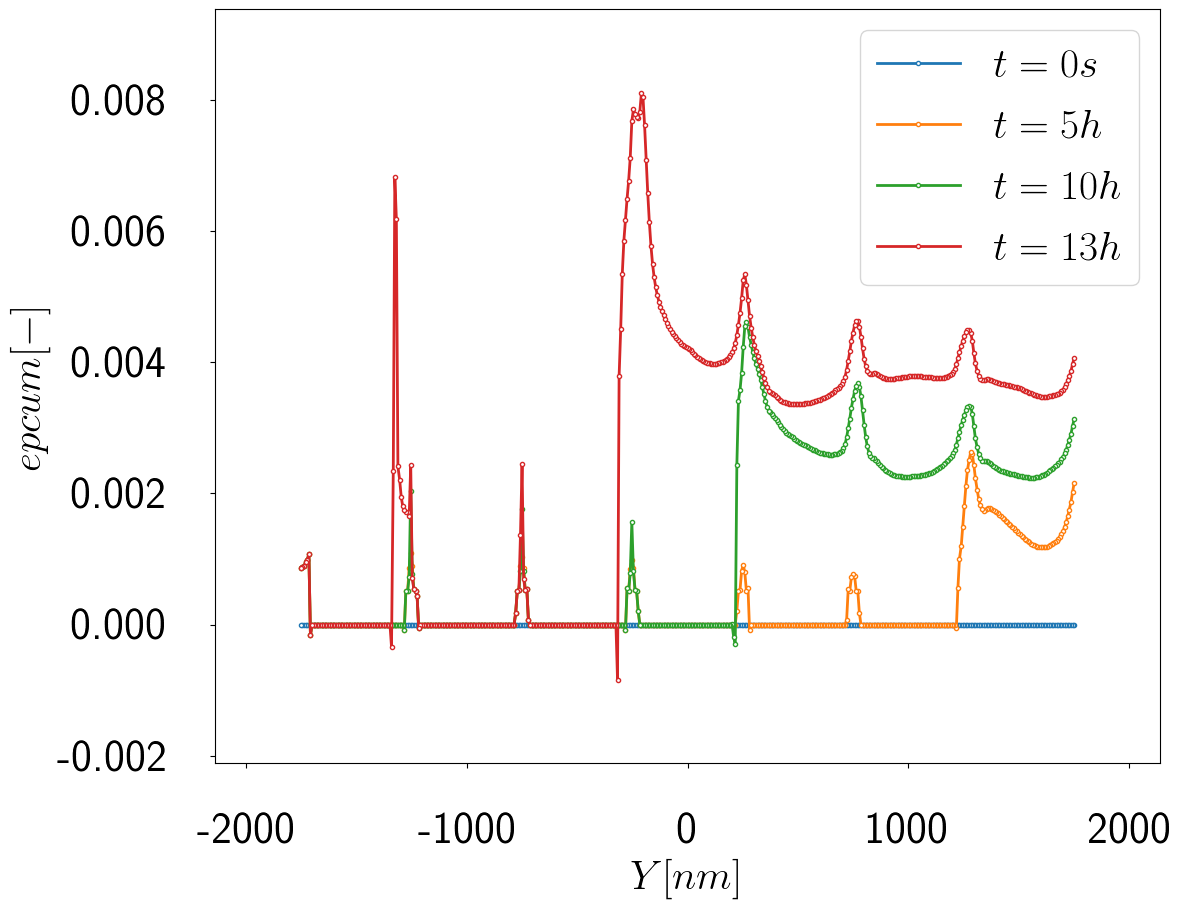}
        \caption{Cumulated plastic strain \label{fig:ppr}}
    \end{subfigure}
    \caption{Outward Al flux  and tensile creep; Evolution with time along the line crossing \rev{the initial } precipitates (continuous line in Figure \ref{fig:JoutwsigCetP})}
    \label{fig:profilsCetP}
\end{figure}

\begin{figure}[ht]
    \centering
    \begin{subfigure}[b]{.8\columnwidth}
        \centering
        \includegraphics[width=\columnwidth]{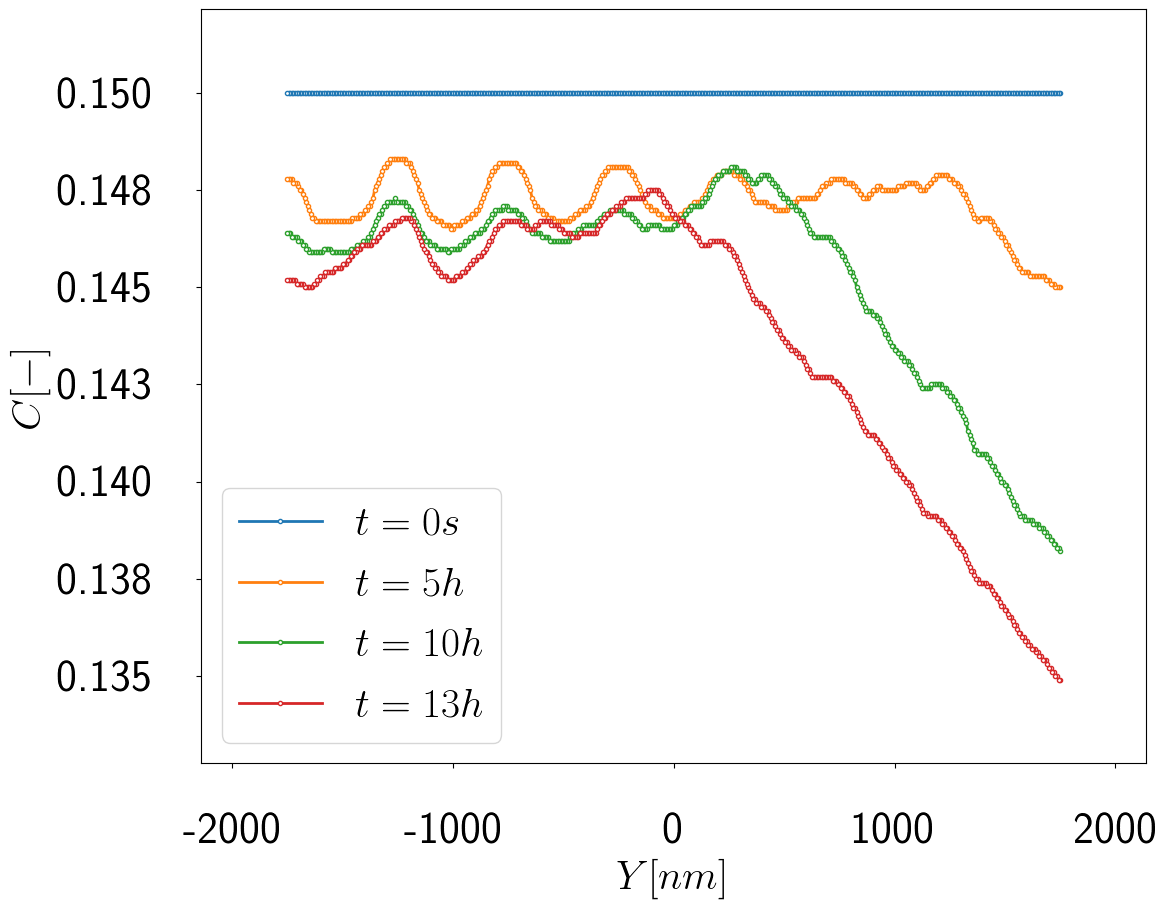}
        \caption{Al concentration \label{fig:Alch}}
    \end{subfigure}
    \begin{subfigure}[b]{.8\columnwidth}
        \centering
        \includegraphics[width=\columnwidth]{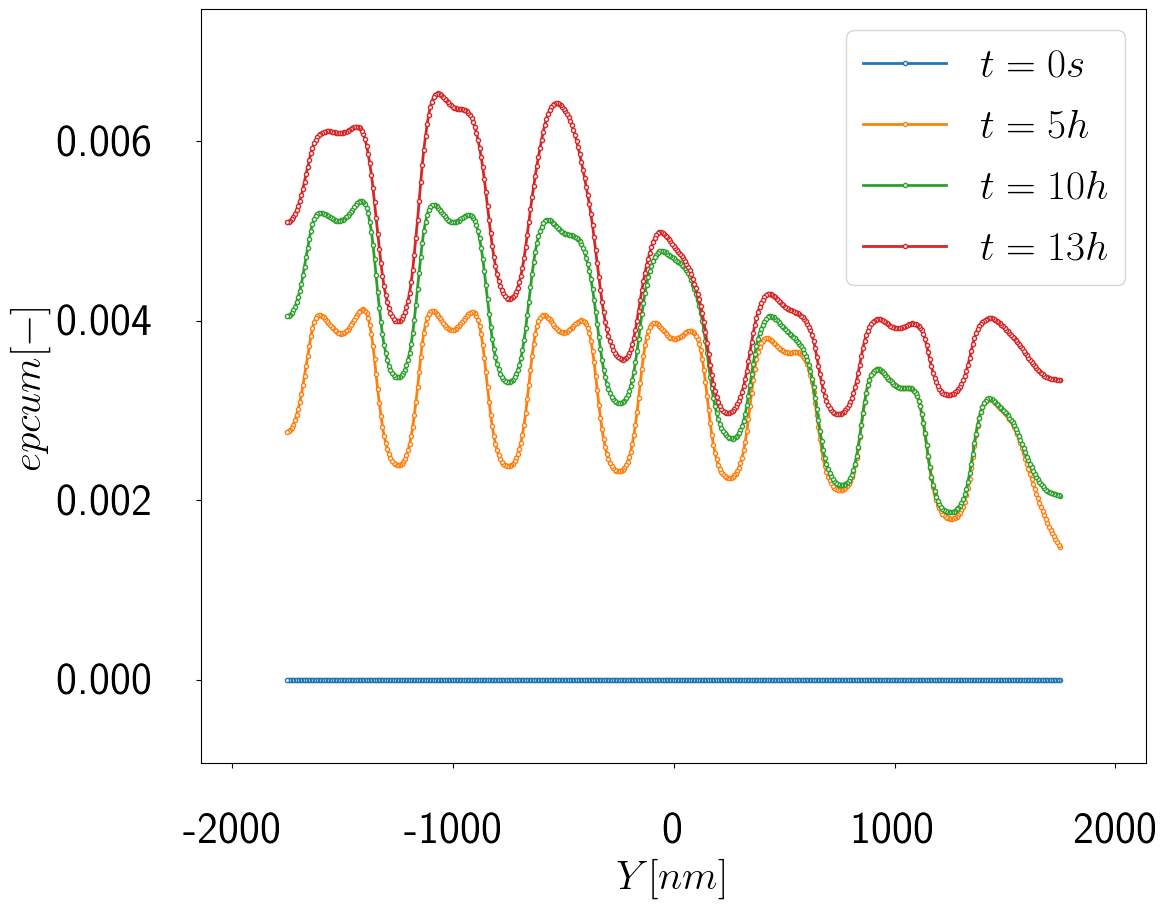}
        \caption{Cumulated plastic strain \label{fig:pch}}
    \end{subfigure}
    \caption{Outward Al flux and tensile creep; Evolution with time along the line between precipitates, \rev{corresponding to the initial $\gamma$ matrix} (dashed line in Figure \ref{fig:JoutwsigCetP})}
    \label{fig:profilsch}
\end{figure}

To get a clearer view of these aspects, we plot the profiles along continuous and dashed lines corresponding to the center of the precipitate row and the $\gamma$ matrix in the center of two rows, respectively, see Figure \ref{fig:JoutwsigCetP}. The corresponding values for different times of exposure to creep and diffusion are shown in Figures \ref{fig:profilsCetP} and \ref{fig:profilsch}. 

The concentration within the precipitates is initially assumed to be homogeneous, but a gradient is observed before and after the precipitates dissolve, Figure \ref{fig:profilsCetP}(a). Moreover, within precipitates, a edge effect is observed on the opposite side of the prescribed flux, where a local enrichment in Al is systematically obtained. On the other hand, after dissolution of the precipitates, despite local variations of the concentration (local minima at the initial $\gamma$ channel position), the gradient is global and seems to be prescribed only by the outward Al flux. Interestingly, local maxima of Al content in precipitates correspond to adjacent local maxima of strain in the $\gamma$ channel, and local maxima of plasticity correspond strictly to the location of minima of Al content, Figure \ref{fig:profilsCetP}(b). Looking at the initial $\gamma$ channel, the gradient driven by the outward Al flux is again evident, in addition to the local oscillations observed in the Al content at a lower range of composition, Figure \ref{fig:profilsch}(a). The associated plasticity is as follows: maxima of Al content correspond to minima of plasticity and vice versa, minima of Al content correspond to maxima of plasticity, Figure \ref{fig:profilsch}(b).

It should be noted that initially, in the presence of $\gamma'$ precipitates, no plasticity can occur due to the assumption of elastic behavior of the precipitates. In short, the above detailed analysis of the gradients shows that the plastic strain localization largely controls the local Al flux.  This point explains the correlation between the narrow shape of the precipitates in the presence of strain localization observed for figure \ref{fig:JoutwsigCetP}.

\subsection{Global Behavior}
\begin{figure}[ht] 		\centering
	   \includegraphics[width=0.8\columnwidth]{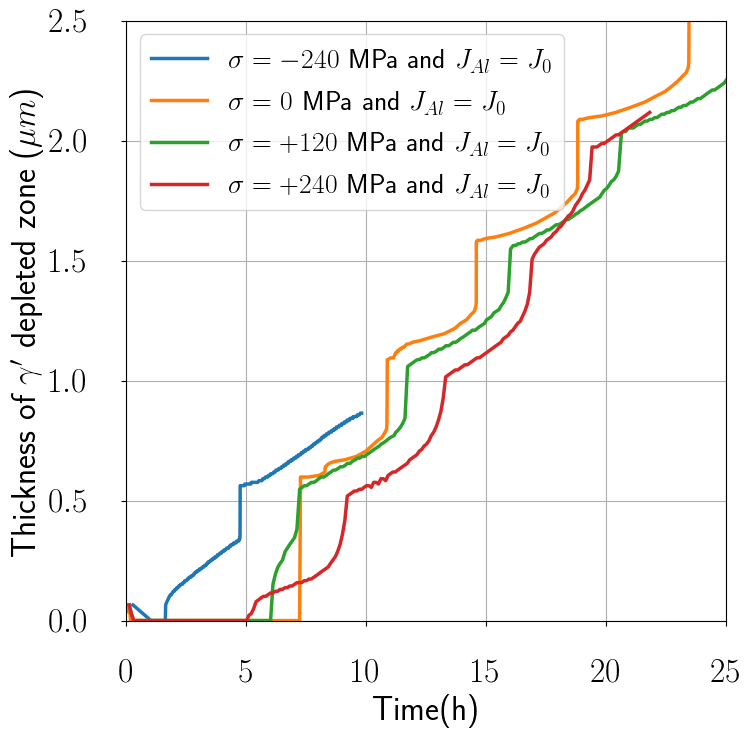}
	\caption{Thickness of the depleted zone in $\gamma'$ precipitates considering the first row of initial precipitates, \rev{for various stress condition and same outward AL flux set to J=J$_0$}}
	\label{fig:depletion}
\end{figure}

In order to clarify the mechanical/diffusion coupling at the length scale  of the chosen model (i.e. the geometry of 7x4 precipitates), we analyze here the sensitivity of $\gamma'$ dissolution to outward flux intensity and applied stress level, Figure \ref{fig:depletion}. This figure is obtained by measuring the size of the $\gamma$ channel extension corresponding to the set of $\gamma'$ precipitates with the highest dissolution rate function of the initial variant of $\gamma'$ (denoted by the continuous line in Figure \ref{fig:JoutwsigCetP}). The outward flux is set to the value $J_0$, varying only the applied stress level and sign. 



First, the dissolution rate is greatly increased by compressive creep compared to tensile creep, see Figure \ref{fig:depletion}. 

In contrast, for positive applied stress and rafts orthogonal to the boundary of outward Al flux, the stress and strain localization actions are more mitigated by the way an intermediate stress level could decrease the dissolution rate: the depletion rate is lower for \SI{120}{MPa} applied in tension, than for no applied stress, see green and orange curves respectively in Figure \ref{fig:depletion}, and for maximum applied stress, the dissolution is less prononced than for other conditions after 10~h even though the dissolution rate seems to accelerate for longer exposure time, see red curve in Figure \ref{fig:depletion}.

\begin{figure}[ht] 		\centering
	   \includegraphics[width=0.8\columnwidth]{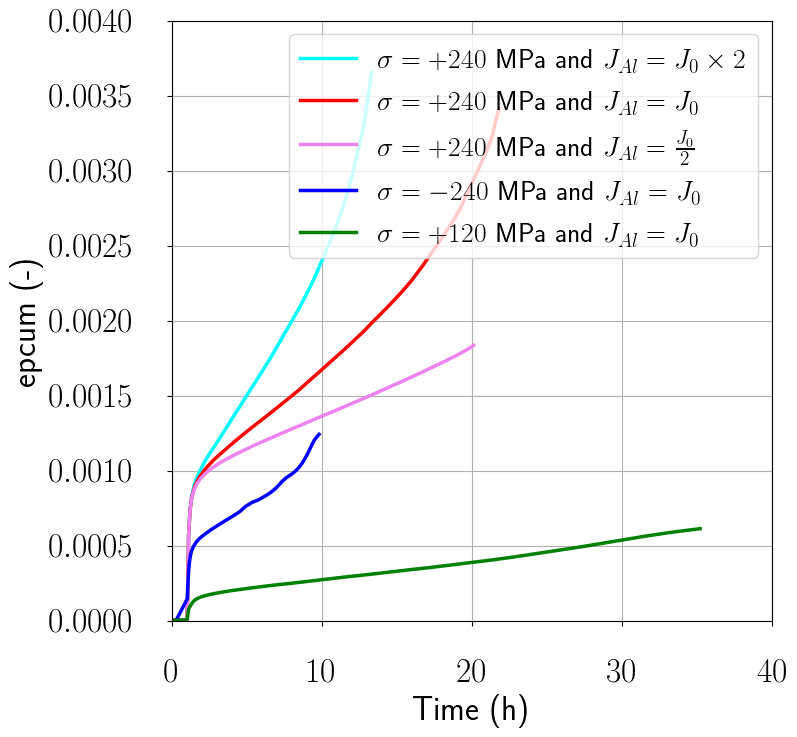}
	\caption{Cumulated plastic strain for various stress and level of outward Al flux}
	\label{fig:epcumver}
\end{figure}

The consequence of these microstructural evolutions is the  ability of the material to resist creep. Therefore, the average plastic strain in the whole model has been plotted as a function of time varying stress level and outward flux of Al, Figure \ref{fig:epcumver}. While the differences in terms of $\gamma'$ depletion were rather small, it is apparent that the variation of the positive stress level has a drastic effect for a given outward flux of Al, compare the green and red curves corresponding to $\sigma=120$ and \SI{240}{MPa} respectively for $J=J_0$. In addition, by setting the applied stress level to $\sigma$=\SI{240}{MPa}, but varying the outward flux of Al from $J_0/2$ to $2\times J_0$, the effective creep rate was also drastically increased. The case of compression, despite the higher dissolution rate of $\gamma'$, leads to initially lower plasticity, see blue curves in Figures \ref{fig:depletion} and \ref{fig:epcumver}, respectively (the maximum time for this case was more limited than for other cases due to the drastic decrease in the computational time increment).

This analysis sheds a light on the strong coupling between diffusion and mechanical coupling: on the one hand, the mechanical behavior is strongly modified by slight variations of the microstructure, and on the other hand, these curves exhibit pseudo-tertiary creep without any ingredient of damage modeling. This last point proves that standard approaches with uncoupled model or weak coupling between microstructure evolution and mechanical properties could integrate in their identification procedures a strong coupling through phase field analysis as developed in the proposed study.

Finally, the issue of damage and microstructure on effective mechanical properties should be carefully analyzed by in-depth comparison with experimental results, both in terms of the size of the phase field model, which should definitely be increased to gain in validity range of the analysis, and in terms of more realistic loading in terms of applied stress levels up to anisothermal loading conditions.

\section{Conclusion}
This study summarizes several aspects of mechanical-diffusion coupling in the context of coated superalloys. Experiments highlight the drastic effect of diffusion on rafting, with emphasis on the role of both the coated and free surfaces. Surprisingly, the competition of rafting with the applied stress level may, to some extent, be of less influence than this diffusional effect. In a nutschell, this study aims to precise the strong stress to diffusion coupling, through experimental and numerical sensitivity analysis using the powerful tool offered by the phase field \rev{finite element } model implementation to account for complex boundary conditions. As an important result, it should be noted that local plasticity governs the local Al flux as a direct consequence of the coupling between diffusion and mechanical state: strain localization acts as a new flux of Al and flux of Al drives the plasticity. The sensitivity analysis shows that even small differences in the microstructure can drastically affect the global mechanical behavior in terms of creep resistance.
 %


\begin{thebibliography}{10}
	\expandafter\ifx\csname url\endcsname\relax
	\def\url#1{\texttt{#1}}\fi
	\expandafter\ifx\csname urlprefix\endcsname\relax\def\urlprefix{URL }\fi
	\expandafter\ifx\csname href\endcsname\relax
	\def\href#1#2{#2} \def\path#1{#1}\fi
	
	\bibitem{reed2008}
	R.~C. Reed, The superalloys: fundamentals and applications, Cambridge
	university press, 2008.
	
	\bibitem{khan1985superalliages}
	T.~Khan, P.~Caron, D.~Fournier, K.~Harris, Superalliages monocristallins pour
	aubes de turbines-caracterisation et optimisation de l’alliage cmsx-2,
	Materiaux \& Techniques 73~(10-11) (1985) 567--578.
	
	\bibitem{murakumo2004creep}
	T.~Murakumo, T.~Kobayashi, Y.~Koizumi, H.~Harada, Creep behaviour of ni-base
	single-crystal superalloys with various $\gamma$' volume fraction, Acta
	Materialia 52~(12) (2004) 3737--3744.
	
	\bibitem{mughrabi2009microstructural}
	H.~Mughrabi, Microstructural aspects of high temperature deformation of
	monocrystalline nickel base superalloys: some open problems, Materials
	Science and Technology 25~(2) (2009) 191--204.
	
	\bibitem{carry1978apparent}
	C.~Carry, J.~Strudel, Apparent and effective creep parameters in single
	crystals of a nickel base superalloy—ii. secondary creep, Acta metallurgica
	26~(5) (1978) 859--870.
	
	\bibitem{pollock1992creep}
	T.~Pollock, A.~Argon, Creep resistance of cmsx-3 nickel base superalloy single
	crystals, Acta Metallurgica et Materialia 40~(1) (1992) 1--30.
	
	\bibitem{cormier2010very}
	J.~Cormier, M.~Jouiad, F.~Hamon, P.~Villechaise, X.~Milhet, Very high
	temperature creep behavior of a single crystal ni-based superalloy under
	complex thermal cycling conditions, Philosophical Magazine Letters 90~(8)
	(2010) 611--620.
	
	\bibitem{naze2021nickel}
	L.~Naz{\'e}, V.~Maurel, G.~Eggeler, J.~Cormier, G.~Cailletaud, Nickel base
	single crystals across length scales, Elsevier, 2021.
	
	\bibitem{tien1971effect}
	J.~Tien, S.~Copley, The effect of uniaxial stress on the periodic morphology of
	coherent gamma prime precipitates in nickel-base superalloy crystals,
	Metallurgical transactions 2 (1971) 215--219.
	
	\bibitem{ignat1993microstructures}
	M.~Ignat, J.-Y. Buffiere, J.~Chaix, Microstructures induced by a stress
	gradient in a nickel-based superalloy, Acta metallurgica et materialia 41~(3)
	(1993) 855--862.
	
	\bibitem{chang2018micromechanics}
	H.-J. Chang, M.~C. Fivel, J.-L. Strudel, Micromechanics of primary creep in ni
	base superalloys, International Journal of Plasticity 108 (2018) 21--39.
	
	\bibitem{fredholm1984creep}
	A.~Fredholm, J.~Strudel, On the creep resistance of some nickel base single
	crystals, Superalloys 1984 (1984) 211--220.
	
	\bibitem{pandey1984environmental}
	M.~Pandey, B.~Dyson, D.~Taplin, Environmental, stress-state and section-size
	synergisms during creep, Proceedings of the Royal Society of London. A.
	Mathematical and Physical Sciences 393~(1804) (1984) 117--131.
	
	\bibitem{cassenti2009effect}
	B.~Cassenti, A.~Staroselsky, The effect of thickness on the creep response of
	thin-wall single crystal components, Materials Science and Engineering: A
	508~(1-2) (2009) 183--189.
	
	\bibitem{bensch2013influence}
	M.~Bensch, C.~Konrad, E.~Fleischmann, C.~M. Rae, U.~Glatzel, Influence of
	oxidation on near-surface $\gamma$' fraction and resulting creep behaviour of
	single crystal ni-base superalloy m247lc sx, Materials Science and
	Engineering: A 577 (2013) 179--188.
	
	\bibitem{levi2012environmental}
	C.~G. Levi, J.~W. Hutchinson, M.-H. Vidal-S{\'e}tif, C.~A. Johnson,
	Environmental degradation of thermal-barrier coatings by molten deposits, MRS
	bulletin 37~(10) (2012) 932--941.
	
	\bibitem{Remy:1993}
	L.~Remy, High temperature fatigue of superalloys, in: NI, Nickel base alloys;
	SP, Superalloys; NI, Nickel base alloys; SP, Superalloys; NI, Nickel base
	alloys; SP, Superalloys; NI, Nickel base alloys; SP, Superalloys; NI, Nickel
	base alloys; SP, Superal, CNRS, Engineering Materials Advisory Services Ltd,
	339 Halesowen Rd , Cradley Heath, Warley, West Midlands B64 6PH, United
	Kingdom, 1993, pp. 825--834.
	
	\bibitem{nutzel2008damage}
	R.~N{\"u}tzel, E.~Affeldt, M.~G{\"o}ken, Damage evolution during
	thermo-mechanical fatigue of a coated monocrystalline nickel-base superalloy,
	International Journal of Fatigue 30~(2) (2008) 313--317.
	
	\bibitem{Sallot:2015}
	P.~Sallot, V.~Maurel, L.~Remy, F.~N'Guyen, A.~Longuet, Microstructure evolution
	of a platinum-modified nickel-aluminide coating during thermal and
	thermo-mechanical fatigue, Metallurgical and Materials Transactions A 46~(10)
	(2015) 4589--4600.
	\newblock \href {https://doi.org/10.1007/s11661-015-2857-9}
	{\path{doi:10.1007/s11661-015-2857-9}}.
	
	\bibitem{kirka2015influence}
	M.~Kirka, K.~Brindley, R.~Neu, S.~Antolovich, S.~Shinde, P.~Gravett, Influence
	of coarsened and rafted microstructures on the thermomechanical fatigue of a
	ni-base superalloy, International Journal of Fatigue 81 (2015) 191--201.
	
	\bibitem{ai2023thermomechanical}
	X.~Ai, L.~Shi, F.~Luo, H.~Pei, Z.~Wen, Thermomechanical fatigue of nickel-based
	single-crystal superalloys, Engineering Fracture Mechanics 284 (2023) 109262.
	
	\bibitem{liu2022coating}
	Y.~Liu, H.~Zhou, M.~Wu, H.~Duan, Y.~Ru, Y.~Pei, S.~Li, S.~Gong, H.~Zhang,
	Coating-related deterioration mechanism of creep performance at a thermal
	exposed single crystal ni-base superalloy, Materials Characterization 187
	(2022) 111839.
	
	\bibitem{zhao2023thickness}
	H.~Zhao, W.~Guo, W.~Zhao, Y.~Ru, J.~Wang, Y.~Pei, S.~Gong, S.~Li, Thickness
	effects on oxidation behavior and consequent $\gamma$’degradation of a
	high-al ni-based single crystal superalloy, Crystals 13~(2) (2023) 234.
	
	\bibitem{lv2022stress}
	J.~Lv, Y.~Zhao, S.~Wang, X.~Zhao, J.~Zhao, L.~Zheng, Y.~Guo, G.~Schmitz, B.~Ge,
	Stress state mechanism of thickness debit effect in creep performances of a
	ni-based single crystal superalloy, International Journal of Plasticity 159
	(2022) 103470.
	
	\bibitem{mataveli2018thin}
	L.~Mataveli~Suave, A.~S. Mu{\~n}oz, A.~Gaubert, G.~Benoit, L.~Marcin,
	P.~Kontis, P.~Villechaise, J.~Cormier, Thin-wall debit in creep of ds200+ hf
	alloy, Metallurgical and Materials Transactions A 49 (2018) 4012--4028.
	
	\bibitem{wang2019combined}
	C.~Wang, M.~A. Ali, S.~Gao, J.~V. Goerler, I.~Steinbach, Combined phase-field
	crystal plasticity simulation of p-and n-type rafting in co-based
	superalloys, Acta Materialia 175 (2019) 21--34.
	
	\bibitem{yu2020thickness}
	Z.~Yu, X.~Wang, H.~Liang, Z.~Li, L.~Li, Z.~Yue, Thickness debit effect in
	ni-based single crystal superalloys at different stress levels, International
	Journal of Mechanical Sciences 170 (2020) 105357.
	
	\bibitem{Mevrel:1987}
	R.~Mevrel, R.~Pichoir, Les revetements par diffusion, Materials Science and
	Engineering 88 (1987) 1 -- 9, proceedings of the First International
	Symposium on High Temperature Corrosion of Materials and Coatings for Energy
	Systems and Turboengines.
	\newblock \href {https://doi.org/https://doi.org/10.1016/0025-5416(87)90060-7}
	{\path{doi:https://doi.org/10.1016/0025-5416(87)90060-7}}.
	
	\bibitem{Koster:1994}
	A.~Koster, E.~Fleury, E.~Vasseur, L.~Remy, Thermal-mechanical fatigue testing.
	in: Automation in fatigue and fracture: Testing and analysis (edited by c.
	amzallag), ASTM STP 1231 (1994) 563--580.
	
	\bibitem{Maurel:2010a}
	V.~Maurel, A.~Koster, L.~Remy, An analysis of thermal gradient impact in
	thermal mechanical fatigue testing, Fatigue Fract. Engng. Mater. Struct.
	33~(8) (2010) 473--489.
	
	\bibitem{cottura2012phase}
	M.~Cottura, Y.~Le~Bouar, A.~Finel, B.~Appolaire, K.~Ammar, S.~Forest, A phase
	field model incorporating strain gradient viscoplasticity: application to
	rafting in ni-base superalloys, Journal of the Mechanics and Physics of
	Solids 60~(7) (2012) 1243--1256.
	
	\bibitem{cottura2016coupling}
	M.~Cottura, B.~Appolaire, A.~Finel, Y.~Le~Bouar, Coupling the phase field
	method for diffusive transformations with dislocation density-based crystal
	plasticity: Application to ni-based superalloys, Journal of the Mechanics and
	Physics of Solids 94 (2016) 473--489.
	
\end{thebibliography}

\end{document}